\documentclass[prl,10pt,twocolumn,amsmath,amssymb,showpacs,floatfix]{revtex4}

\usepackage{times}
\usepackage{graphicx}
\usepackage{dcolumn}
\usepackage{amsmath}
\usepackage{amssymb}

\newcommand{\dd}{\mathrm{d}}

\begin{document}
\title{Downfolding electron-phonon Hamiltonians from ab-initio calculations: 
application to K$_3$Picene}

\author{Gianluca Giovannetti$^{1}$, Michele Casula,$^{2}$ Philipp
  Werner$^{3}$, Francesco Mauri$^{2}$, and Massimo
  Capone$^{1}$}
\address{$^1$CNR-IOM-Democritos National Simulation Centre and International School
for Advanced Studies (SISSA), Via Bonomea 265, I-34136, Trieste, Italy \\
$^2$CNRS and Institut de Min\'eralogie et de Physique des Milieux condens\'es,
Universit\'e Pierre et Marie Curie, 4 place Jussieu, 75252 Paris,
France \\
$^3$Department of Physics, University of Fribourg, 1700 Fribourg, Switzerland
}
\date{\today}
\begin{abstract} 
We propose an electron-phonon parameterization which reliably
reproduces the geometry and harmonic frequencies of a real system. 
With respect to standard electron-phonon models, it adds a
``double-counting'' correction, which takes into account the lattice
deformation as the system is 
dressed by low-energy electron-phonon processes. We show the importance of this correction by
studying
potassium-doped picene
(K$_3$Picene), recently claimed to be a superconductor with
a $T_c$ 
of up to 18 K. The Hamiltonian parameters are derived 
from ab-initio density functional theory, and the lattice model is
solved by dynamical mean-field theory.
Our calculations include
the effects of electron-electron interactions and local
electron-phonon couplings. 
Even with the inclusion of a strongly coupled molecular phonon, the
Hubbard repulsion prevails and the system is an insulator with 
a small Mott gap of $\approx$ 0.2 eV.
\end{abstract}

\maketitle

{\it Introduction}
In recent years, a significant effort has been made to derive
low-energy Hamiltonians from ab-initio electronic structure
calculations in order to model the effect of strong electron
correlations in a predictive
fashion \cite{kotliar-review,casula_screening}. 
Despite
remarkable progress 
in the field, little attention has been paid on how to
include lattice vibrations coupled to
electrons in those Hamiltonians, with coupling strengths
taken from experiments or from first principles.
The major difficulties are to correctly estimate the ``bare'' couplings, i.e. the
ones undressed from electron-electron (EE) or electron-phonon (EP) scattering
processes explicitly treated in the low-energy
manifold \cite{crpa,arita_gRPA}, and to 
avoid ``double counting (DC)'', i.e. summing up effects
already treated in the model Hamiltonians. 

Dealing with both EE and EP
interactions is particularly important in molecular crystals, which
are characterized  by a tight competition between interactions.
Phonons are thought to drive superconductivity close to the Mott regime, 
with unconventional features \cite{CaponeRMP,CaponeScience}. 
Superconductivity has indeed been found in the
fullerides \cite{GunnarssonRMP}, and later in 
the family of so-called ``aromatic superconductors'', such as 
picene (K$_3$picene) \cite{Mitsuhashi2010}, 
coronene \cite{KubozonoCoronene}, and 
1,2:8,9-dibenzopentacene \cite{Xue2011}, with $T_c$ up to 33 K by intercalation with
alkali atoms. These compounds are appealing from the viewpoint of potential  
applications, but their physics is 
poorly understood. Indeed, the metallicity and superconductivity are
highly debated \cite{lannoo,picene_prl_casula,Subedi,Giovannetti,Kim,Valenti}. One
difficulty is to intercalate large enough 
crystals, and some experimental groups have found an insulating
behavior of K$_3$picene at low temperature \cite{Valenti,Goldoni}.  

A common theoretical framework to study these systems is based on the
Hubbard-Holstein Hamiltonian, where the electrons experience local
interactions and are coupled to local
vibrations. We use a generalized
Holstein model where the local molecular mode does not simply
couple with the charge density, and off-diagonal couplings in the
orbital basis are included. We show that a proper derivation of the EP matrix elements must include a
DC correction to counteract the effect of the lattice 
relaxation already included in the low-energy Hamiltonian.
We provide a prescription to compute the EP-DC
correction from experimental or ab-initio estimates of the geometry
and phonon frequencies.  We demonstrate the importance of this term,
applying our theory to build a low-energy Hamiltonian for
K$_3$Picene with intramolecular Hubbard interactions and EP couplings
whose amplitudes have been determined by density functional theory
(DFT) calculations. 
We show that despite the strength of the EP coupling, the Coulomb
repulsion prevails and the system is an insulator with a small gap of
0.2 eV.

{\it Theory} 
Let us start from the tight-binding Hamiltonian
$H_\textrm{tb}$, which gives the low-energy
band structure:
\begin{equation}
H_\textrm{tb}=\sum_{\alpha\beta\sigma ij} t^{\alpha \beta}_{ij} c^\dagger_{\alpha
  \sigma i}
c_{\beta \sigma j} - \mu \sum_{\alpha\sigma i} n_{\alpha \sigma i} ,
\label{tight-binding}
\end{equation}
where $c^\dagger_{\alpha \sigma i}$ ($c_{\alpha \sigma i}$) creates (annihilates) an electron 
on the lattice site $i$ with spin $\sigma$ in the orbital
$\alpha$, and $n_{\alpha \sigma i} = c^\dagger_{\alpha \sigma
  i} c_{\alpha \sigma i}$.
The model in
Eq.~(\ref{tight-binding}) is usually derived from a DFT electronic
structure computed for the lattice geometry \emph{relaxed} at a given
chemical potential $\mu$ 
(system at rest).

In the first step, let us  consider classical phonons. 
We are going to quantize them later. We parameterize the
EP coupling in the system as a single-mode Holstein
phonon of frequency $\omega_\textrm{bare}$,
locally coupled to the electronic manifold of
Eq.~(\ref{tight-binding}) via the $\delta V^\textrm{bare}_{\alpha \beta}$ \cite{vdef} matrix elements: 
\begin{equation}
H_\textrm{el-ph} = H_\textrm{tb} + \sum_{\alpha \beta \sigma i} r_i
\delta V_{\alpha
  \beta}^\textrm{bare} c^\dagger_{\alpha \sigma i} c_{\beta \sigma i} +
\frac{\omega^2_\textrm{bare}}{2} \sum_i  \left ( r_i - r^0 \right )^2,
\label{el-phon_classical}
\end{equation}
with $r_i$ classical phonon displacements and
the shift corresponding to the structural minimum 
parametrized 
by
$r^0$ , which is in general non-zero and gives rise to non-trivial
effects, as we show below.
The bare couplings are such that the model solution at
the given filling yields the equilibrium geometry ($r_i=0$) and the phonon
frequency ($\omega_\textrm{dressed}$) of the physical system at the
same filling.
In other words:
\begin{eqnarray}
&&\left. \frac{\partial \langle  H_\textrm{el-ph} \rangle}{\partial r_i} \right|_{r_i=0} =  0 
\label{geometry}, \quad 
\left. \frac{\partial^2 \langle  H_\textrm{el-ph} \rangle}{\partial r_i^2} \right|_{r_i=0} = \omega^2_\textrm{dressed},
\label{frequency}
\end{eqnarray}
which states that the force vanishes at the equilibrium position
and the harmonic contribution to the ion displacement is given by
$\omega_\textrm{dressed}$. 

By quantizing the phonon in Eq.~(\ref{el-phon_classical}) we obtain:
\begin{eqnarray}
H_\textrm{el-ph} &  = &  H_\textrm{tb} + \sum_{\alpha \beta \sigma i} \sqrt{2} x^0
g_{\alpha \beta}^\textrm{bare} c^\dagger_{\alpha \sigma i} c_{\beta
  \sigma i}  \nonumber \\
& + &
\sum_{\alpha \beta \sigma i} (a_i + a^\dagger_i) g_{\alpha \beta}^\textrm{bare} c^\dagger_{
  \alpha \sigma i} c_{\beta \sigma i} +  \omega_\textrm{bare} \sum_i a^\dagger_i a_i, 
\label{el-phon}
\end{eqnarray}
where now $g^\textrm{bare}_{\alpha \beta} = \delta V^\textrm{bare}
_{\alpha\beta} / \sqrt{2 \omega_\textrm{bare}}$, $x_i =
\sqrt{\omega_\textrm{bare}} (r_i - r^0) = \langle
a_i + a^\dagger_i \rangle/\sqrt{2} $ is the dimensionless displacement,
and $x^0=\sqrt{\omega_\textrm{bare}} r^0$. In Eq.~(\ref{el-phon}),
the $\sum_{\alpha \beta \sigma i} \sqrt{2} x^0  g_{\alpha \beta}^\textrm{bare}
c^\dagger_{\alpha \sigma i} c_{\beta \sigma i} $ term appears as a correction to the
usual EP Hamiltonian. A closer inspection reveals that
this term yields a band deformation related to the modified geometry
before filling the low-energy bands. This correction is
necessary as the tight-binding model is defined at the given filling, while the bare quantities are computed by undressing the
system from the low-energy electrons. Analogously, the
renormalization of the phonon frequency from $\omega_\textrm{bare}$ to
$\omega_\textrm{dressed}$ is due to the EP interaction
acting on the low-energy manifold.

Note that the spirit of including the EP-DC correction provided by
$x^0$ is the same as for the DC correction of the electronic part, necessary whenever an
EE interaction is explicitly 
added 
in Eq.~(\ref{tight-binding}). 
Usually, we require the EE-DC correction to provide the original DFT band structure when the many-body system
is solved at the mean-field level.  
Analogously, if we take the bare $\delta V_{\alpha \beta}$ 
as the \emph{variation} of the interaction due to the phonon
displacement, we adjust the EP-DC term such that the
mean-field solution of the model corresponds to the ab-initio band
structure, geometry and phonon frequency. 
This gives a prescription on how to evaluate the EP term. Once 
$\delta V^\textrm{bare}_{\alpha \beta}$ is computed from ab-initio
calculations, $\omega_\textrm{bare}$ and $x^0$ are set by solving
Eqs.~(\ref{geometry}) self-consistently at the mean-field level. This is also doable in 
the 
presence of EE
interactions.  We now apply our theory to K$_3$picene and show
the importance of including the EP-DC term to get meaningful results.

{\it Model} To study and reproduce the properties of K$_3$Picene, we 
choose the following low-energy Hamiltonian:
\begin{eqnarray}
H &=& H_\textrm{el-ph} + \epsilon_\textrm{EE-DC} \sum_{\alpha \sigma i} n_{\alpha \sigma i}
+ U \sum_{\alpha \sigma i}  n_{\alpha \sigma i} n_{\alpha -\sigma i} 
\nonumber \\ 
& + & U^\prime \sum_{\substack{ \alpha \beta \sigma i \\ \alpha \ne \beta}} n_{\alpha \sigma i} n_{\beta -\sigma i} + 
 (U^\prime - J)\sum_{\substack{\alpha \beta \sigma i \\ \alpha \ne \beta}}  n_{\alpha \sigma i}  n_{\beta \sigma i},
\label{hamiltonian}
\end{eqnarray}
where we add the EE part to the $H_\textrm{el-ph}$ in
Eq.~(\ref{el-phon}), parameterized through $U$, $U^\prime$, and $J$
Hubbard and Hund parameters. 
In this case, the EE-DC correction $\epsilon_\textrm{EE-DC}$ is just 
a redefinition of the chemical potential shift $\mu$.

The parameters in Eq.~(\ref{hamiltonian}) are obtained by ab-initio DFT calculations 
within the local density approximation (LDA) carried out with the Quantum Espresso \cite{QE} package. 
The lattice unit cell has been taken from powder diffraction data \cite{Mitsuhashi2010}, 
and the internal molecular coordinates relaxed 
by energy minimization \cite{picene_prl_casula}. The hoppings $t_{ij}^{\alpha\beta}$ of the tight binding model are derived 
from the Wannier construction \cite{wannier90}, in order to reproduce
the LDA low-energy bands $\epsilon_{m \sigma}(\textbf{k})$.
The maximally localized Wannier functions (MLWFs) are built by choosing an energy window which includes bands 
originating from the three lowest unoccupied molecular orbitals (LUMO, LUMO+1 and LUMO+2) of the neutral molecule \cite{notaWFS}.
The localized orbital set of Eq.~(\ref{hamiltonian}) is defined by a rotation of the MLWF basis such that  
the local $H_\textrm{tb}=-t_{ii}^{\alpha\beta}$ is \emph{diagonal}. This corresponds to 
working 
with molecular orbitals (MOs) 
which are very close to the actual MOs of an isolated picene molecule,
as explained in Ref.~\onlinecite{picene_prb_casula}.

The local EE interaction in Eq.~(\ref{hamiltonian}) is justified 
by the molecular nature of the crystal, with the on-site (molecular) repulsion 
larger than the other energy scales of the system. According to constrained-random phase approximation (c-RPA) estimates \cite{Nomura},
the nearest-neighbor repulsion is around 1/3 of 
the local $U$. The values of the full
local interaction matrix have been computed in Ref.~\onlinecite{Nomura} by the c-RPA method in the two-orbital MLWF basis.
We obtain the corresponding interaction in the MO basis by rotation, which gives $U$=0.68 eV, $U^\prime$=0.63 eV, $J$=0.10 eV.
We extend these values to the three-MO model of Eq.~(\ref{hamiltonian}), by assuming that they are insensitive to 
the MO type, and by neglecting de-screening due to the LUMO+2 channel. However, this is a minor effect 
compared to the large screening coming from the full frequency dependence of $U(\omega)$, 
which goes up to 4.4 eV in the unscreened ($\omega \rightarrow \infty$) limit ($U_\textrm{bare}$). 
In Ref.~\onlinecite{casula_screening}, we have proven that the correct low-energy model which 
includes the high-energy screening processes is 
the Hamiltonian with the $U(\omega=0)$ static interaction and the 
bandwidth $t$ renormalized by the factor 
$Z_B =  \exp \left( 1/\pi  \int_0^\infty \!\!\! \dd\omega  ~  \textrm{Im} U(\omega) /\omega^2 \right).$
We estimate $Z_B$ from the experimental loss function ($\textrm{Im}[-1/\epsilon(\omega)]$) of K$_3$Picene, which has been
measured up to 40 eV by electron energy-loss spectroscopy \cite{Roth}. By neglecting the crystal momentum
dependence of the full dielectric function (much smaller than its energy dependence in a molecular crystal),
we can obtain a rough estimate of the imaginary part of the retarded $U$ as 
$\textrm{Im}U(\omega) \approx  U_\textrm{bare}
\textrm{Im}[-1/\epsilon(\omega)]$. Using a low-energy cutoff
corresponding to the MOs included in the model we obtain a
renormalization $Z_B$ = 0.76 for all the hoppings in Eq.~(\ref{hamiltonian}).

\begin{figure}[ht]
\includegraphics[width=1.0\columnwidth]{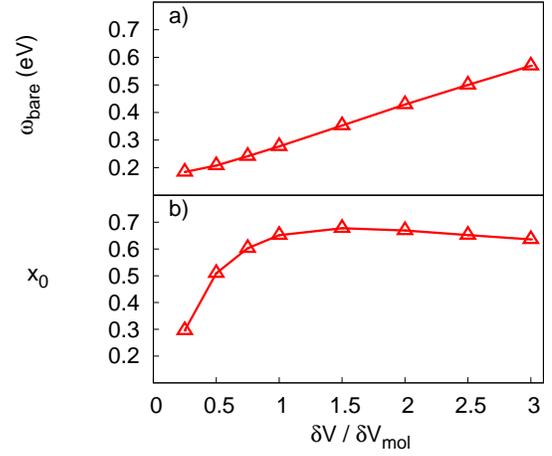}
\caption{Mean-field solution of Eqs.~(\ref{geometry}) 
  as a function of the deformation potential $\delta
  V$, taken with respect to the ab-initio molecular value $\delta
  V_\textrm{mol}$. The geometry is constrained to the crystal relaxed DFT
  solution $x_\textrm{min}=0$, and the 
  dressed frequency is set to the crystal ab-initio DFT value of
  $\omega_\textrm{dressed}=0.173$ eV, yielding
  $\omega_\textrm{bare}$ and $x^0$ reported in panel (a) and (b), respectively.
}
\label{elph_corr}
\end{figure}

To make the many-body calculations feasible, we parameterize the 
phonon branches $\omega_{\textbf{q}\nu}$ ($\textbf{q}$ is the phonon momentum and $\nu$ is the phonon mode)  
by a single monochromatic local (molecular) phonon. We take the
molecular phonon with the largest $|g^\textrm{bare}|$ as the 
representative of the total EP coupling.
The $g$ matrix is screened by both EE and
EP processes within the low-energy manifold. To undress the system
from screening involving LUMO+$n$ states and obtain the \emph{bare} couplings, we
perform a density functional perturbation theory (DFPT) calculation \cite{dfpt,QE} 
of a neutral isolated molecule, instead of undressing
$g$ by c-RPA as recently suggested in Ref.~\onlinecite{arita_gRPA}. 
Therefore $\delta V^\textrm{bare}_{\alpha\beta} \approx \delta
V^\textrm{mol}_{\alpha\beta}$, and the
local bare matrix elements are the molecular ones.
The molecular phonon frequency of the most coupled mode is 0.193 eV, and
its corresponding phonon frequency in the crystal is
$\omega_\textrm{dressed}=0.173$ eV (from DFPT calculations of the crystal).
From $\delta V^\textrm{mol}_{\alpha\beta}$ computed by DFTP and from
$\omega_\textrm{bare}=0.277$ eV evaluated through
Eqs.~(\ref{geometry})
(Fig.~\ref{elph_corr}), we get
$g^\textrm{bare}_{\alpha\beta}$ (in eV):
\begin{equation}
\left( \begin{array}{ccc}
~~~0.066 & -0.010  & -0.002 \\ 
-0.010 & -0.038  & -0.051 \\
-0.002 & -0.051  & -0.018  \end{array} \right).
\end{equation}
Note that $g$ has sizable inter-orbital matrix elements, of the same
magnitude as the diagonal ones. 
In the following, we are going to study the dependence 
of the solution on the EP coupling strength by taking into
account 3 sets of $g^\textrm{bare}_{\alpha\beta}$, based on $\delta
V^\text{mol}_{\alpha \beta}$, $2~\delta V^\text{mol}_{\alpha \beta}$, and $3~\delta
V^\text{mol}_{\alpha \beta}$.

{\it Methods} In order to solve the Hamiltonian in Eq.~(\ref{hamiltonian}) with the above
parameters, we use dynamical mean field theory (DMFT)
\cite{antoine_rmp}. The DMFT equations are solved with an
exact-diagonalization (ED) impurity solver \cite{Caffarel}, and some
of the results are cross-checked using a continuous time quantum Monte
Carlo (CTQMC) solver \cite{Werner2006}.   
DMFT maps the Hubbard-Holstein lattice problem (\ref{hamiltonian})
onto an Anderson-Holstein impurity model (AHIM) \cite{Hewson} with a
self-consistently defined bath. 
In order to solve this impurity problem using the Lanczos method we
have to truncate the Hilbert space by limiting the number of bath
levels $N_\text{bath}$ (here 9) and the maximum number of excited phonons (here
we use 20). We have therefore a
three-orbital impurity whose 
local non-interacting Hamiltonian 
contains the EP-DC correction
$h_{\alpha\beta} = (E^{\alpha}-\mu)\delta_{\alpha\beta}+
\sqrt{2} x^0g_{\alpha \beta}^\textrm{bare} $. The impurity is
hybridized with $N_\text{bath}$ bath levels with energy
${\epsilon}_l$, and coupled 
to 
an on-site harmonic oscillator. 
$c_{\alpha\sigma}$ denotes the annihilation operator for the impurity
level $\alpha$ with spin $\sigma$, $b_{l\sigma}$ the operator for
the $l$th level in the bath and $a$ the operator for a local phonon of frequency $\omega_0$:
\begin{align}
H_\text{AHIM}=& \sum_{{\alpha\beta}{\sigma}} h_{\alpha\beta} c^{\dagger}_{\alpha\sigma}c_{\beta\sigma}  + U\sum_{{\alpha}{\sigma}}n_{{\alpha}{\sigma}} n_{{\alpha}-{\sigma}} \nonumber\\
&+ U'\sum_{{\alpha} {\beta}{\sigma}\atop \alpha\ne\beta}n_{{\alpha}{\sigma}} n_{{\beta}-{\sigma}} + (U'-J)\sum_{{\alpha}{\beta}{\sigma}\atop \alpha\ne \beta}n_{{\alpha}{\sigma}} n_{{\beta}{\sigma}} \nonumber\\ 
& + \sum_{l{\sigma}} {\epsilon}_l b^\dagger_{l{\sigma}}b_{l{\sigma}}  + \sum_{l{\alpha}{\sigma}} V_{l{\alpha}} ( c^\dagger_{{\alpha}{\sigma}} b_{l{\sigma}} + h.c.) \nonumber\\
& + \sum_{\alpha\beta\sigma} g_{\alpha\beta} c^\dagger_{{\alpha}{\sigma}}c_{{\beta}{\sigma}} (a^\dagger + a) + \omega_0 a^\dagger a.
\label{hahim}
\end{align}
Then the dynamical Weiss field 
${{\mathcal{G}}_0}^{-1}_{{\alpha}{\beta}}$ can be defined as 
\begin{equation}
{{\mathcal{G}}_0}^{-1}_{{\alpha}{\beta}} (i {\omega}_n)=i {\omega}_n -h_{\alpha\beta}
- \sum_{l=1}^{N_\text{bath}} \frac{ V^*_{l\alpha} {V}_{l\beta}  }{i {\omega}_n - {\epsilon}_l }.
\label{weiss}
\end{equation}

\begin{figure}[ht]
\includegraphics[width=.75\columnwidth,angle=0]{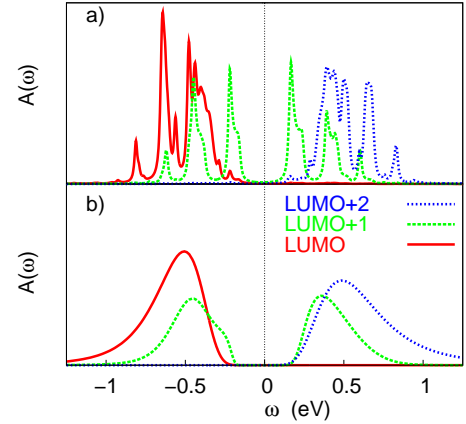}
\caption{(Color online) Paramagnetic spectral functions obtained by ED
  (a) and CTQMC (b) considering c-RPA EE interactions
  and a bandwidth reduction $Z_B$ at $T=0$K (ED) and $\beta = 1/(k_BT)$=100 eV$^{-1}$
  (CTQMC).} 
\label{fig1}
\end{figure}

Note that the bath function in Eq.~(\ref{weiss}) has off-diagonal
components. Correspondingly, we 
have to compute all elements of the impurity Green's function matrix
G$_{{\alpha}{\beta}}$, and the self-energy $\Sigma_{{\alpha}{\beta}}$
will also have off-diagonal components. The local
lattice Green's function is
$G_\text{loc}^{{\alpha}{\beta}} (i {\omega}_n) = \sum_{k} (i {\omega}_n + \mu -H^\text{DFT}_{{\alpha}{\beta}}(k)-{\Sigma}_{{\alpha}{\beta}} )^{-1}$,
where the sum runs over the Brillouin zone and
$H^\text{DFT}_{{\alpha}{\beta}}(k)$ is the Fourier transform of the DFT-LDA
non-interacting Hamiltonian. 
By equating $G_{{\alpha}{\beta}}$ to $G_\text{loc}^{{\alpha}{\beta}}$ 
we can obtain a {\it new}
Weiss field which is then fitted to Eq.~(\ref{weiss}) and determines
the new set of parameters ${V}_{l\alpha}$ and ${\epsilon}_l$. 
The above procedure is iterated until convergence is reached. 

Since non-diagonal EP terms cannot be treated with the
Monte Carlo technique of Ref.~\cite{Werner2007}, we restrict the CTQMC
calculations to the model without EP coupling. In the MO basis it
turns out that the sign problem is negligible, even though the
off-diagonal hybridizations are relatively large.  

{\it Results} By taking the electronic part of our Hamiltonian
(\ref{hamiltonian}) only, 
we find K$_3$Picene to be a Mott insulator: the
LUMO (LUMO+2) orbital is completely filled 
(empty) while the orbital LUMO+1 is half-filled and has well
pronounced Hubbard bands hybridized with the LUMO and LUMO+2 orbitals
(see Fig.~\ref{fig1}). 
This insulating state is consistent with the result of previous DMFT
calculations (in which a much larger $U$ was used)
\cite{Valenti}. However, in our case the Mott gap is significantly
smaller (gap half-width of $\approx$ 
0.2 eV) and the system is quite close to the Mott transition.
We note that the results obtained using the ED and CTQMC solvers are
consistent, so this finding is not affected by 
truncations 
in the
ED treatment. 

The discrepancy between the Mott insulating behavior of K$_3$Picene
found here and the recent reports of superconducting signatures 
suggests an important role of EP interactions in stabilizing the
superconducting phase. 
We thus add 
in our ED/DMFT scheme the
Holstein-type $g_{\alpha\beta}$ terms. 
We first discuss the results without EP-DC correction.
In this case the effect of the EP interaction on the 
electronic structure is remarkable. Table~\ref{occupation_table} lists the
MO occupations found in the ED/DMFT solution of
Eq.~(\ref{hamiltonian}) with EP coupling strengths of
different magnitude. 
The coupling with phonons moves the LUMO+1 orbital
away from half-filling, and induces a strong hybridization
between the LUMO and LUMO+2 orbitals, which increases as we increase
the EP coupling.

\begin{table}[b]
\caption{MO occupations from the ED/DMFT solution of
  Eq.~(\ref{hamiltonian}). Results
  are reported for a system with and without EP
  double-counting terms in the lower and upper 
  panel, respectively, 
  for different
  EP couplings. The purely electronic case
  ($\delta V =0$) is also shown. The right column is the most populated
  phonon level $N_\text{ph}^\text{max}$.
}
\label{occupation_table}
\begin{ruledtabular}
\begin{tabular}{ l | d | d | d | d}
\makebox[50pt][c]{} & \makebox[22pt][l]{LUMO} &
\makebox[22pt][l]{LUMO+1} & \makebox[22pt][l]{LUMO+2} &
\makebox[5pt][c]{$N_\text{ph}^\text{max}$}\\
\hline
\makebox[50pt][l]{$\delta V=0$} & 1.00  & 0.50 & 0.00  & - \\
\hline
\multicolumn{5}{c}{Without EP-DC correction
  ($\omega_\textrm{bare}=0.193$ eV, $x^0=0$)} \\
\hline
\makebox[50pt][l]{$\delta V = \delta V_\textrm{mol}$} & 1.00 & 0.45 & 0.05  &
1\\  
\makebox[50pt][l]{$\delta V = 2\delta V_\textrm{mol}$} & 1.00 & 0.29 & 0.21  &
3 \\
\makebox[50pt][l]{$\delta V = 3\delta V_\textrm{mol}$} & 1.00 & 0.25 & 0.25  &
9 \\
\hline
\multicolumn{5}{c}{With EP-DC correction
  ($\omega_\textrm{bare}$ and $x^0$ from Fig.~\ref{elph_corr}) } \\
\hline
\makebox[50pt][l]{$\delta V = \delta V_\textrm{mol}$} & 0.99 & 0.50 & 0.01  &
0 \\ 
\makebox[50pt][l]{$\delta V = 2\delta V_\textrm{mol}$} & 0.98 & 0.50 & 0.02  &
0\\
\makebox[50pt][l]{$\delta V = 3\delta V_\textrm{mol}$} & 0.94 & 0.53 & 0.03  &
0\\
 \end{tabular}
\end{ruledtabular}
\end{table}

\begin{figure}[!ht]
\includegraphics[width=1.1\columnwidth]{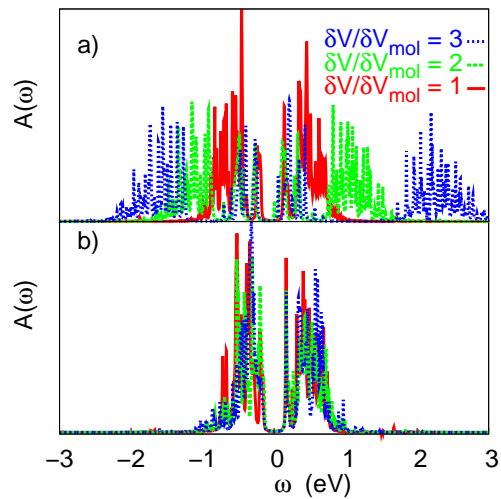}
\caption{(Color online) Panel (a): spectral functions obtained by ED/DMFT
  considering c-RPA electron interactions, a bandwidth reduction $Z_B = 0.76$
 and Holstein-type couplings at $T=0$ without EP-DC correction. Panel
 (b): same as panel (a), but with
 the EP-DC correction term included in the Hamiltonian.}
\label{fig3}
\end{figure}

For $\delta V = 3\delta V_\textrm{mol}$
both LUMO+1 and LUMO+2 are 1/4 filled, and the system is prone to
metallicity (although still on the insulating side).
To understand the origin of this effect, we analyze the phonon
population distribution.  It features a broad peak centered around
the $n=9$ phonon level, a Frank-Condon behavior related to the
molecular deformation. The system geometry changes as the
EP coupling increases, by pulling the minimum away from the original
center of the phonon oscillators. This is 
clear from the last column of Tab.~\ref{occupation_table}, where the
phonon peak shifts to higher levels as the coupling gets stronger.
This has several consequences: it mixes the unperturbed MO's
states already at the molecular (on-site) level and
drifts the bands leading to a more asymmetric structure and to the
observed occupations (see Fig.~\ref{fig3}(a)). These effects are related
to the off-diagonal EP couplings which transfer electrons between
orbitals. These terms are resilient to the Hubbard interaction as
opposed to the density terms which are quenched by strong correlations\cite{Sangiovanni}.

The result changes both qualitatively and quantitatively when the
EP-DC correction is added.
The deformation
driven by the bare EP coupling is counterbalanced by the
EP-DC correction, which constrains the model to have the correct
ab-initio DFT geometry when it is solved at the mean-field level. The
$\omega_\textrm{bare}$ and $x^0$ 
fixed by that constraint are plotted in Fig.~\ref{elph_corr}, as a
function of $\delta V$. We find that $\omega_\textrm{bare}$ increases
linearly with $\delta V$, while $x^0$ saturates after a first linear growth. 
The ED/DMFT spectrum of the model with EP-DC correction is shown in
Fig.~\ref{fig3}(b). The effect of phonons is much less dramatic. The
spectrum and electron populations remain close to the results in the
absence of EP coupling,
while the phonon population stays peaked at the lowest phonon state
(Tab.~\ref{occupation_table}), signaling that with the EP-DC correction
the EE correlation alone is not able to deform the DFT
geometry. 
The system remains a small gap insulator, as in the case without phonons.

In conclusion,
we have shown the importance of including the EP
double-counting correction to model the EP coupling from
ab-initio DFT results or experimental data. K$_3$Picene is an
ideal test-case for our theory, as in molecular crystals the bandwidth,
the local EP coupling
and the local EE repulsions live on the same energy scale,
and the properties result from a subtle competition between
them. Therefore, theoretical predictions are extremely sensitive to the
quality of the model. Using the LDA+DMFT approach, we found that 
K$_3$Picene is a Mott insulator. The local Hubbard repulsion opens a small gap of $\approx$
0.2 eV, while local Holstein phonons, whose coupling has been
estimated from ab-initio molecular calculations, do not modify
qualitatively the electronic structure when the EP-DC correction is added.
The results are qualitatively modified by the EP-DC correction, which calls
for a critical reanalysis of the theoretical work done so far on
EP models for molecular crystals, whenever the EP-DC correction has
been disregarded.
Our EP parameterization goes well beyond the
specific case of  K$_3$Picene, and can be applied also to more general EP Hamiltonians.

We thank A. Amaricci, H. Aoki, R. Arita and T. Kariyado for fruitful
and stimulating discussions.
Massimo Capone and GG acknowledge financial support by the European
Union through FP7/ERC Starting Independent Research Grant ``SUPERBAD"
(Grant Agreement n. 240524) and of FP7-NMP- 2011-EU-Japan project
LEMSUPER (Grant Agreement no. 283214). Computing Time has been provided
by CINECA through project CONDMAG (lsB04). Michele Casula thanks the IDRIS-GENCI for computing
time under project 96493. PW acknowledges support from SNF Grant 200021\_140648 and computing
time on the Brutus cluster at ETH Zurich.


\begin{thebibliography}{00}
\bibitem{notaWFS} Wannierization has been performed on a $4\times4\times4$
electron-momentum mesh in a window of 0.7 eV around the Fermi level with Wannierized bands in a perfect agreement with those computed in DFT.
\end{thebibliography}
\end{document}